\documentclass{iaus}
\usepackage{graphics}

  \checkfont{eurm10}
  \iffontfound
    \IfFileExists{upmath.sty}
      {\typeout{^^JFound AMS Euler Roman fonts on the system,
                   using the 'upmath' package.^^J}%
       \usepackage{upmath}}
      {\typeout{^^JFound AMS Euler Roman fonts on the system, but you
                   dont seem to have the}%
       \typeout{'upmath' package installed. iaus.cls can take advantage
                 of these fonts,^^Jif you use 'upmath' package.^^J}%
      }
  \else
  \fi

  \checkfont{msam10}
  \iffontfound
    \IfFileExists{amssymb.sty}
      {\typeout{^^JFound AMS Symbol fonts on the system, using the
                'amssymb' package.^^J}%
       \usepackage{amssymb}%
       \let\le=\leqslant  
         
      }{}
  \fi

  \IfFileExists{amsbsy.sty}
    {\typeout{^^JFound the 'amsbsy' package on the system, using it.^^J}%
     \usepackage{amsbsy}}
    {\providecommand\boldsymbol[1]{\mbox{\boldmath $##1$}}}

  \usepackage{psfig}

\def\gsim{\mathrel{\raise0.35ex\hbox{$\scriptstyle >$}\kern-0.6em 
\lower0.40ex\hbox{{$\scriptstyle \sim$}}}}
\def\lsim{\mathrel{\raise0.35ex\hbox{$\scriptstyle <$}\kern-0.6em 
\lower0.40ex\hbox{{$\scriptstyle \sim$}}}}

\title[Outskirts of Galaxy Clusters: intense life in the suburbs]
      {Superclusters and Voids in the Sloan DSS}

\author[V. M\"uller {\it et al.\/}]%
{Volker M\"uller$^1$ \and Christian Maulbetsch$^1$}

\affiliation{$^1$Astrophysikalisches Institut, Potsdam 14482, Germany\\
email: vmueller@aip.de}

\pubyear{2004}
\volume{195}
\pagerange{1--3}
\date{?? and in revised form ??}
\jname{Outskirts of Galaxy Clusters: intense life in the suburbs}
\editors{A. Diaferio, ed.}
\begin{document}

\maketitle

\begin{abstract}
We analyze quasi-2-dimensional slices of the SDSS EDR. Gaussian smoothing 
with weighting by the inverse of the selection function provides 2D 
density fields across the full survey depth. Superclusters (SC) are
characterized by a percolation algorithm in the large-scale smoothed field. 
Group candidates are identified with density 
maxima in the small-scale smoothed field. The group mass function depends  
on the SC environmental density. We derive the shape-dependent
3-point correlation function and the void size distribution. These are 
well reproduced by the galaxies identified in high-resolution $\Lambda$CDM 
simulations.

\end{abstract}

\firstsection 

\section{Cosmic Density Fields}

\begin{figure}
\centerline{
\vspace{4.5cm}
{\Large 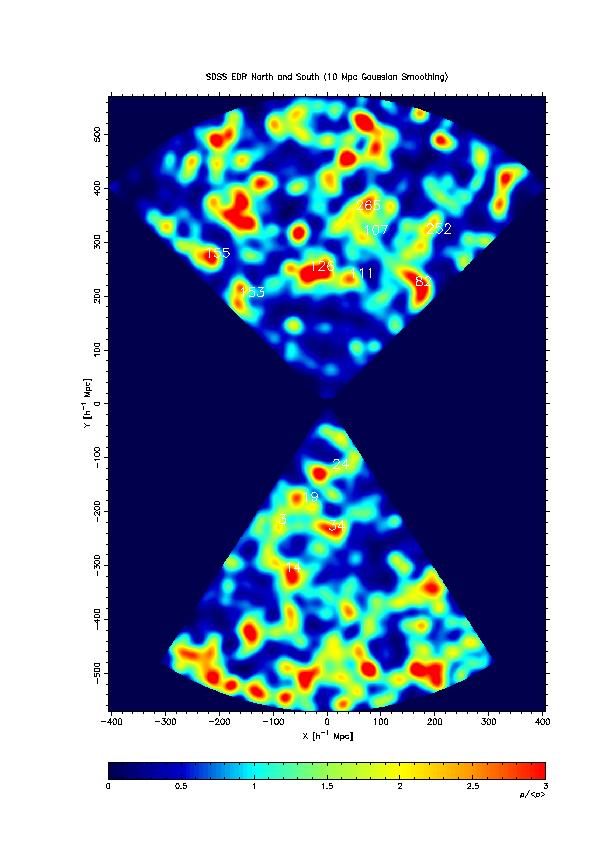 and 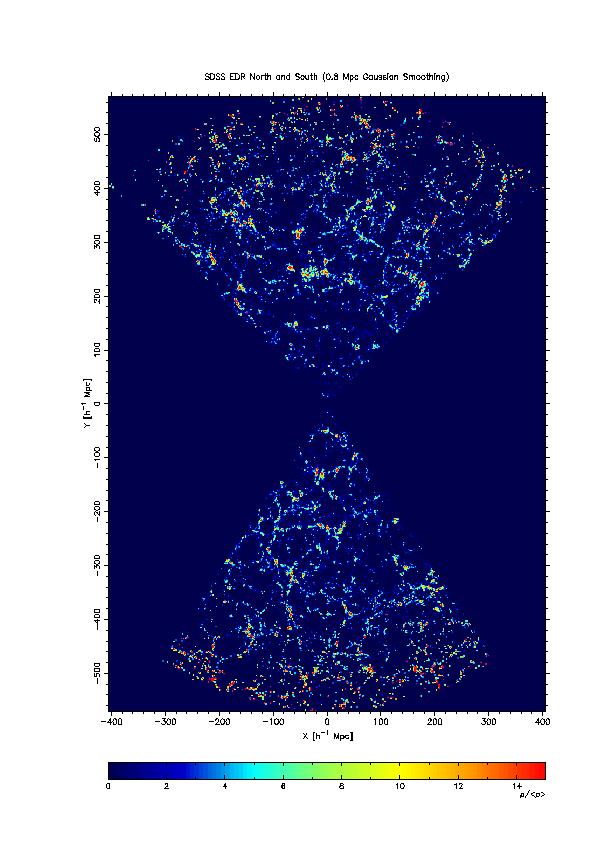}
\vspace{4.5cm}
}
\caption{SDSS EDR density field with 10~$h^{-1}$Mpc\ (left panel) and 
0.8~$h^{-1}$Mpc\ (right panel) smoothing. North is on top. Numbers correspond
to superclusters from \cite[Einasto et al. (2001)]{Einasto01}.}
\label{fig:sdss_sm}
\end{figure}

Establishing large-scale cosmic density fields from the observed galaxy 
distribution was an early aim of analyzing galaxy redshift surveys 
\cite[(Saunders et al. 1991)]{Saunders91}. 
Averaging over local random inhomogeneities, the density field provides
genuine measures of the complex pattern of sheets and filaments in the
universe comprising the `cosmic web' 
\cite[(Bond, Kofman, \& Pogosyan 1996)]{Bond96}. 
Our analysis is based on the ongoing SLOAN Redshift Survey
intended to derive one million galaxy redshifts within one quarter of the sky.
Here we analyze the first preliminary data release of 34 000 galaxies in two
$2.5 ^\circ$ thick stripes along the celestial equator in the Northern (top) 
and Southern (bottom) hemispheres. 
This part of the survey is 95 \% complete for red
magnitudes $ 13.0 \le r^* \le 17.7$. We constrain the analysis to a depth of
$z=0.2$ that corresponds to a limiting distance of 570 $h^{-1}$Mpc~ ($h \approx 0.7$).

The cosmic density field in Figure~\ref{fig:sdss_sm} was Gaussian smoothed
with the 
galaxy selection 
function as inverse weight to get an approximately uniform coverage
of the survey volume \cite[(Einasto et al. 2003)]{Einasto03}. 
We identify 43
superclusters as the largest nonlinear structures in the survey.  The
most massive system in the North is the Supercluster 126 containing 7
Abell clusters and 6 X-ray clusters from the ROSAT bright survey 
\cite[(Schwope et al. 2000)]{Schwope00}. Most superclusters form complex 
multi-branching filaments with one
or two strong central concentrations. The richest system in the South is the
Pegasus-Pisces supercluster 3 with 9 Abell clusters.

The high-resolution density field provides a detailed insight into the
cosmic web. All supercluster concentrations are subdivided into interconnected
filamentary branches with galaxies like pearls on a string. We identify about
5000 galaxy groups, for which we derive harmonic radii and the velocity
dispersions to estimate the virial mass for gravitationally bound systems. We
find that the upper limit of the group masses is about 5 times higher in the
central supercluster regions.  Furthermore, we find that the group masses 
scatter over a larger range in overdense regions due to a further evolution
of the mass hierarchy. 
The environmental dependence of group and cluster properties was established 
in high-resolution numerical simulations of galaxy formation.

\section{The 3-Point Correlation Function and Void Size Distribution}

\begin{figure}
\centerline{
\psfig{file=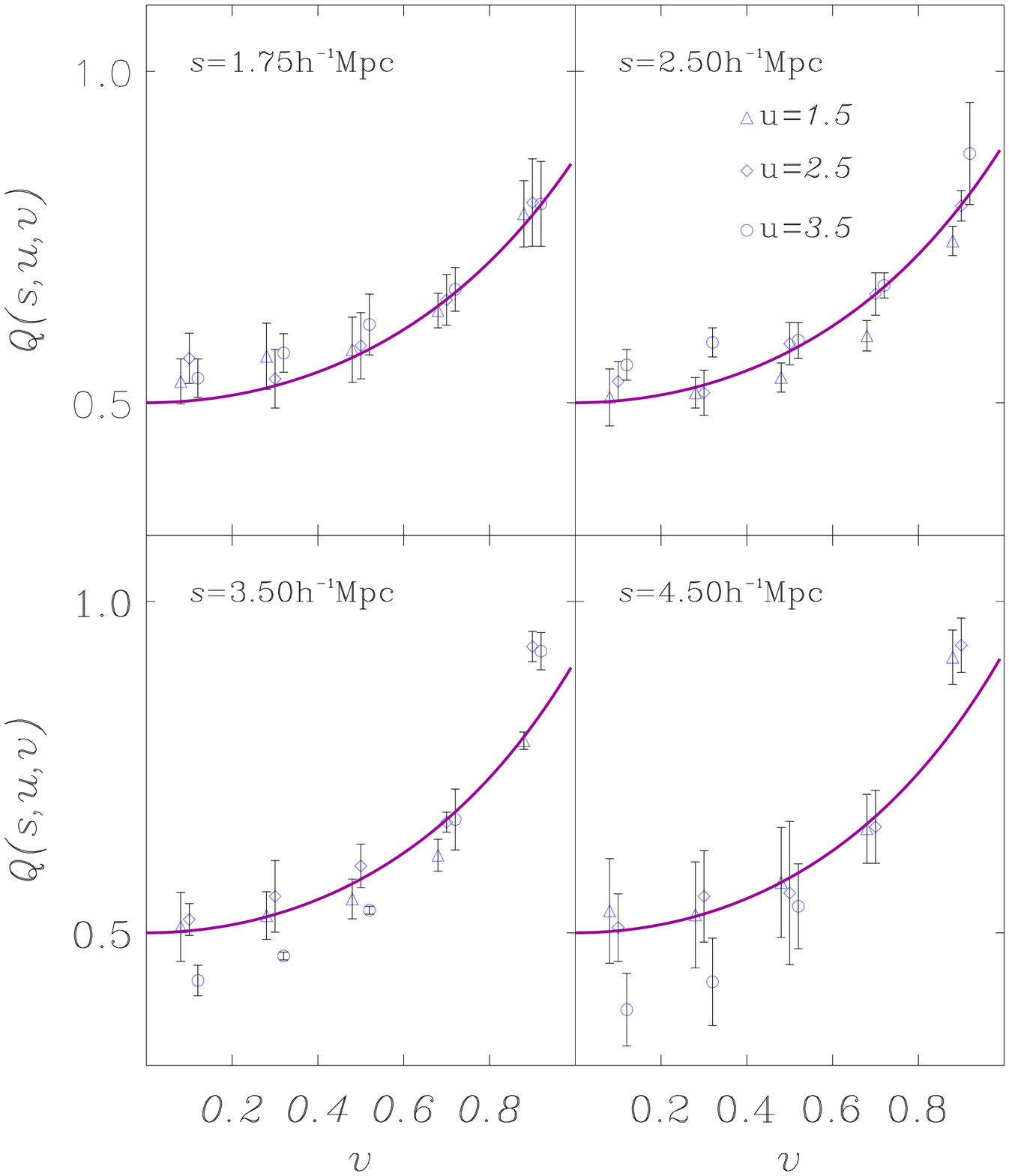,angle=0,width=6.8cm}
\psfig{file=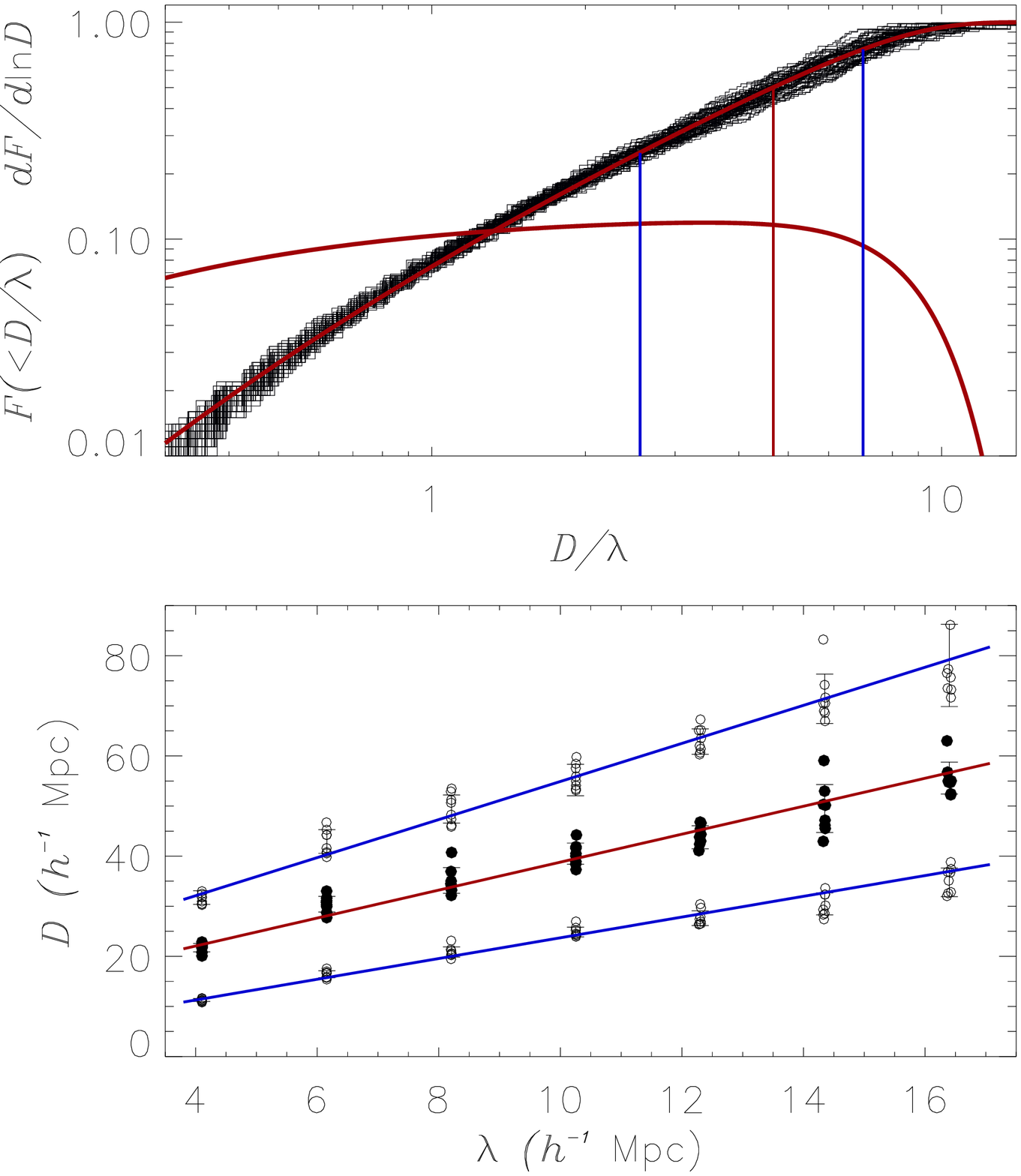,angle=0,width=6.8cm}}
\caption{Left: The reduced redshift space correlation function in the 
strongly clustered regime. Right: Cumulative (histograms) and fitted 
differential (solid line) void size distribution. 
The lower panel shows the scaling of 
median and quartile void sizes for random diluting the data.}
\label{fig:stat}
\end{figure}

We quantify the filamentarity of the galaxy distribution in deriving the
reduced 3-point correlation function $Q$ shown in Figure~\ref{fig:stat}, left panel. 
It is a reliable
measure of hierarchical clustering, sensitive to the morphology of structures
and accessible to gravitational perturbation theory. The four panels show the
redshift space correlation function for galaxies sitting on the edges of
triangles with sides $s$, $us$, and $(u+v)s$ ($u>1$, $0<v<1$). The increase
with $v$ is characteristic for filaments. The solid lines show theoretical
fits derived empirically for LCRS galaxies \cite[(Jing \& B\"orner 1998)]{Jing98}
that are in agreement with our data (cf. also 
\cite[Kayo et al. (2004)]{Kayo04} for a recent analysis of the SDSS first data
release). We compare our results with high-resolution DM-simulations ($256^3$ 
particles with spatial resolution of a few kpc) in a 
60 $h^{-1}$Mpc box size for a concordance $\Lambda$CDM-model. There is a strong
influence of redshift space corrections that remove almost all $u$-dependence
of the reduced 3-point amplitude $Q$ but well preserves the increase of $Q$
with $v$ (anisotropy of filamentarity of the clustering) predicted from
gravitational perturbation theory. We find a weak but significant increase of 
this filamentarity effect for blue versus red galaxies. In the simulations we 
find a stronger anisotropy of $Q$ for halos identified at high redshifts 
($z = 3 \dots 1$) than at the present epoch. Our results indicate that the 
amplitude of the 3-point function $Q$ can be reproduced in high-resolution 
$\Lambda$CDM-simulations and is not overpredicted by a factor of two as claimed 
by \cite[Jing \& B\"orner (1998)]{Jing98}.

We identify a sample of voids covering most of the survey area in seven 
volume-limited data sets from the
SDSS-EDR with the 2D-void finder of 
\cite[Arbabi-Bidgoli \& M\"uller (2002)]{Arbabi02}. The size 
distribution weighted with the void area is shown 
in Figure~\ref{fig:stat} 
as function of the ratio of 
diameter $D$ to the mean galaxy separation $\lambda$, $x=D/\lambda$, 
i.e. ${dF/d\ln D} \propto (x/(x+a))^2 \exp[-(x/a)^4] \mbox{, where } a\approx 9$. 
The broad size distribution with a cut-off is 
in agreement with recent excursion set modeling by  
\cite[Sheth \& Van de Weygaert (2003)]{sheth03}, but it shows no 
abundance peak at that scale. 
The high abundance of small
voids is typical for the galaxy void hierarchy.
Voids among red galaxies (preferentially in clusters)
are 15~\% larger than among blue galaxies. The lower panel of 
Figure~\ref{fig:stat} shows that the void sizes (median and quartiles)  
have a characteristic linear scaling with the mean galaxy 
separation $\lambda$ if the galaxies are randomly diluted. This scaling law 
(cf. \cite[M\"uller et al. (2000)]{mueller00}) characterizes the 
void hierarchy in the universe and in simulations.

\begin{acknowledgments}
We want to thank our collaborators Maret and Jaan Einasto, Douglas Tucker, 
Sepehr Arbabi-Bidgoli, Enn Saar, and Pekka Hein\"am\"aki.
\end{acknowledgments}

\end{document}